\documentclass[a4paper]{iopart}
\usepackage{graphicx}
\usepackage{amsfonts}
\usepackage{iopams}

\newcommand\be{\begin{equation}}
\newcommand\ee{\end{equation}}
\newcommand\bea{\begin{eqnarray}}
\newcommand\eea{\end{eqnarray}}

\newcommand\minus\backslash

\newtheorem{theorem}{Theorem}

\newtheorem{lemma}[theorem]{Lemma}

\begin{document}
\title{Quantum Integrals from Coalgebra Structure}

\author{S Post$^1$ and D Riglioni$^2$  }
 
\address{1 University of Hawai`i, 2656 McCarthy Mall, Honolulu, HI 96822}
\address{2 Centre de Recherches Mat\'ematiques, Universit\'e de Montr\'eal}

\ead{spost@hawaii.edu,  riglioni@crm.umontreal.ca}

\begin{abstract} Quantum versions of the hydrogen atom and the harmonic oscillator are studied on non Euclidean spaces of dimension $N$. $2N-1$ integrals, of arbitrary order,  are constructed via a multi-dimensional version of the factorization method, thus confirming the conjecture of D Riglioni 2013 J. Phys. A: Math. Theor. 46 265207. The systems are extended via coalgebra extension of $\mathfrak{sl}(2)$ representations, although not all integrals are expressible in these generators. As an example, dimensional reduction is applied to 4D systems to obtain extension and new proofs of the superintegrability of known families of Hamiltonians.  
\end{abstract}


\section{Introduction}

A Maximally Superintegrable (MS) system in classical mechanics is an integrable N-dimensional (ND)
Hamiltonian system which is endowed with the maximum possible number of 2N-1 functionally
independent integrals of motion. The study  and classification of superintegrable systems is of central importance on mathematical physics. On the one hand they are a source of exactly solvable models which over the years have found applications in many areas of physics
such as in condensed matter physics as well as atomic, molecular and nuclear physics see
e.g. \cite{ feher1987dynamical, gritsev2001model, quesne2004deformed} and reference therein. On the other hand the symmetry algebra defined by its constants of the motion is of interest in the field of group theory and their representations. The most well-known example of superintegrable systems, the hydrogen atom and the harmonic oscillator, are in correspondence with $\mathfrak{so}(N+1)$. More recently the discovery of superintegrable systems with constants of the motion of arbitrary order in the momentum have been found to be in correspondence with new type of polynomial algebras. Moreover since MS Hamiltonian systems are conjectured to be exactly solvable  their eigenfunctions can be described in terms of either some class of orthogonal polynomials  or for scattering states in terms of some special functions. 
MS systems with quadratic constants of the motion have been completely classified for 2D Riemannian and pseudo-Riemannian spaces\cite{KKMP}. Example of MS systems with constants of the motion of order higher than two are indeed much rarer in literature since a systematic classification of these systems go through the solution of nonlinear differential equations whose complexity increase with the order of the integrals \cite{superreview}.
However some interesting examples of MS systems with constants of the motion of arbitrary high order have been found as a deformation of systems admitting quadratic integrals of motion. Two remarkable examples are given by the so-called TTW \cite{TTW} or PW\cite{PW2010} systems. As was remarked in a recent paper \cite{riglioni2013classical},  the possibility of obtaining higher order MS systems from 2-dim ones can be understood in terms of coalgebra symmetry. For a review of superintegrable systems see \cite{superreview}.

To be self contained, let us recall briefly how to extend systems in higher dimensions by using the coalgebra operators.  We consider the superintegrable extension of the hydrogen atom on a space of constant scalar curvature. The Hamiltonian for the system in 2D is given by  
$$ H=2 k^2 \Delta_{S^2} - \frac{\alpha s_3}{\sqrt{s_1^2+s_2^2}}, \qquad s_1^2+s_2^2+s_3^2=1.$$
Transforming to conformal coordinates via
$$ x_1=\frac{1}{k}\cot(\theta/2)\cos(\phi), \qquad x_2=\frac{1}{k}\cot(\theta/2)\sin(\phi)$$
gives the following radial form of the Hamiltonian
\begin{equation}
\label{eq1}
H = \frac{(1 + k^2 ( x_1^2 + x_2^2) )^2}{2} \left( p_1^2 + p_2^2  \right) - \mu \frac{1 - k^2 ( x_1^2 + x_2^2 )}{\sqrt{x_1^2 + x_2^2}}.
\end{equation}
The trajectories for  bounded motion of  (\ref{eq1}) at regular points in phase space are closed, as an effect of its superintegrability.  

\begin{figure}
\centering
\includegraphics [width=0.6\textwidth]{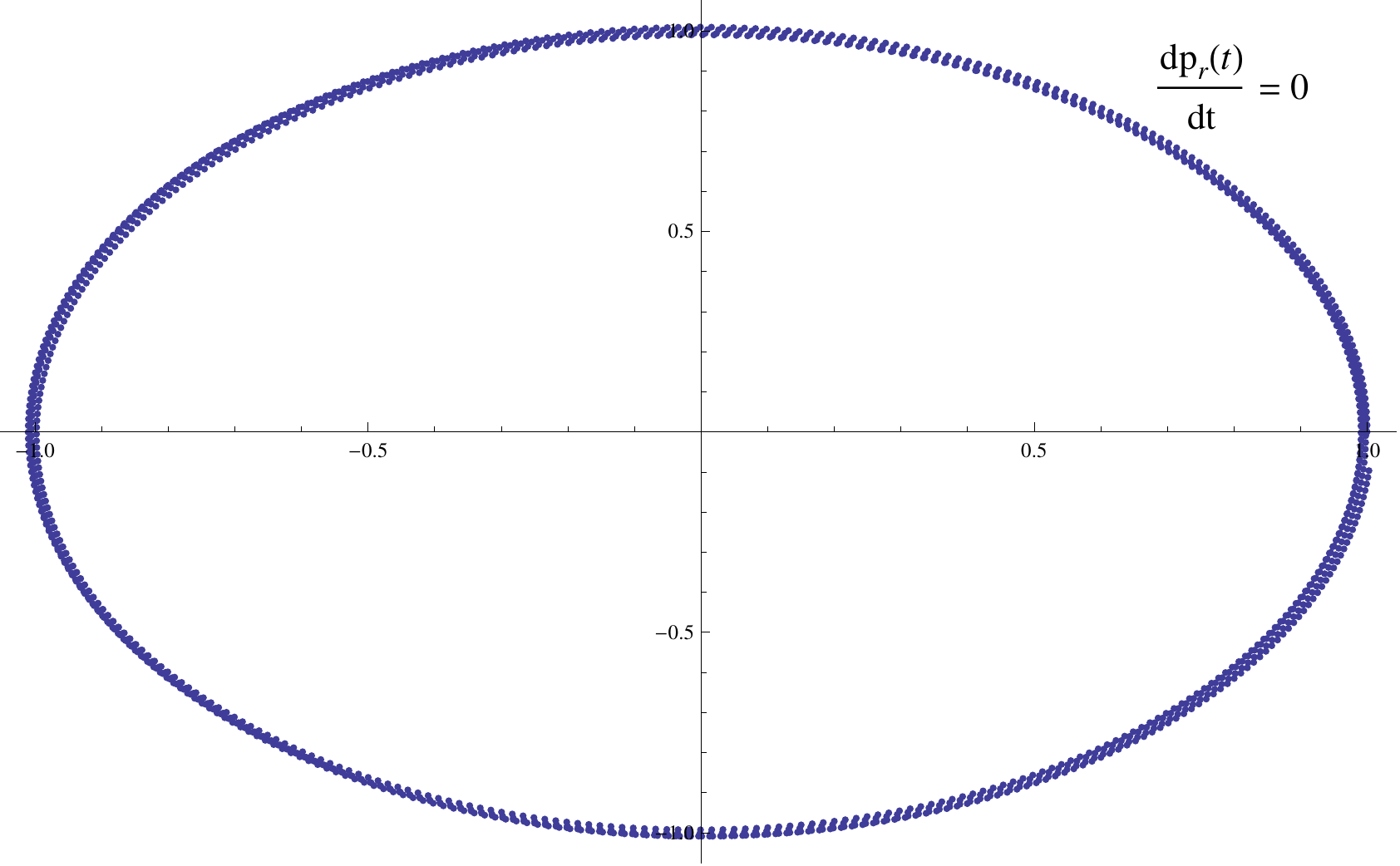}
\caption{$k=-0.04 ; \mu = 60$}
\label{closedtraj} 
\end{figure}

As mentioned above,the system (\ref{eq1}) is characterized by a coalgebra symmetry, namely it is possible to express the system (\ref{eq1}) via a sympletic realization of the Lie algebra $\mathfrak{sl}(2)$
\begin{equation}
\label{rep2}
D_2(J_-) = x_1^2 + x_2^2,    \quad  \quad  D_2 (J_+) = p_{1}^2 + p_{2}^2, \quad D_2(J_3) = x_1 p_{1} + x_2 p_{2},  
\end{equation}
with central element 
\be D_2 (\mathcal{C}) = (x_1 p_{2} - x_2 p_1)^2.
\ee
and (Poisson) bracket
\be  [ f , g ]_{PB} =  \sum_i \partial_i f \partial_{p_{x_i}} g - \partial_i g \partial_{p_{x_i}} f.
\end{equation}
To keep the notation succinct, we drop the subscript for the remainder of the paper.
As a reminder, the generators for the Lie algebra    $\mathfrak{sl}(2)$ satisfy
$$ [J_3, J_+] =2J_+, \qquad [J_3, J_-]=-2J_-, \qquad [J_-, J_+]=4J_3$$
with  Casimir element
$$ C=J_+J_{-}-J_3^2.$$
The representation (\ref{rep2}) can be constructed from a coalgebra of the Poisson realization of $\mathfrak{sl}(2)$; 
namely, the coproduct is given by
\begin{equation*}
\Delta (1) = 1 \otimes 1, \qquad  \Delta(J_i) = J_i \otimes 1 + 1 \otimes J_i,    
\end{equation*}
which preserves the algebra relations
\begin{equation*}
[ \Delta (J_i) , \Delta ( J_j ) ] = \Delta ( [ J_i , J_j ]  ). 
\end{equation*}
Taking repeated coproducts leads to a 2ND phase space realizations for $\mathfrak{sl}(2)$ given by  
\begin{eqnarray*}
J_-^{(N)} \equiv D_N(J_-) = \Delta (D_2(J_-) )^{N-1} & = & \sum_{i=1}^N x_i^2 \\
J_+^{(N)} \equiv  D_N(J_+) = \Delta (D_2(J_+) )^{N-1} & = & \sum_{i=1}^N p_i^2 \\
J_3^{(N)} \equiv D_N(J_3) = \Delta (D_2(J_3) )^{N-1} &  = & \sum_{i=1}^N x_i p_i.
\end{eqnarray*} 

The main point is that it is possible to express (\ref{eq1}) as:
\begin{eqnarray}
H &\equiv& \frac{(1 + k^2 J_-^{(2)})^2}{2} J_+^{(2)} - \mu \frac{1 - k^2 J_-^{(2)}}{\sqrt{J_-^{(2)}}}\nonumber\\
& =& \frac{(1 + k^2 J_-^{(2)})^2}{2} \left( \frac{(J_3^{(2)})^2}{J_-^{(2)}} + \frac{{C^{(2)}}}{J_-^{(2)}} \right) - \mu \frac{1 - k^2 J_-^{(2)}}{\sqrt{J_-^{(2)}}},   \nonumber
\end{eqnarray}
and hence, using the coproduct, the  higher dimensional realization of the Hamiltonian system can be generalized in a manner preserving its integrability properties. In particular, each of the intermediate Casimir elements will commute with the generators 
\begin{equation*}
[J_k^{(N)} , {C}^{(m)} ] =  0, \qquad m \leq n ; k=+,-,3,
\end{equation*}
Therefore any 2-dimensional system $H(J_-^{(2)} , J_3^{(2)} ,J_+^{(2)} )$ can be generalized to the ND system $H(J_-^{(N)} , J_3^{(N)} , J_+^{(N)} )$ which will have by construction $N$ constants of the  motion given by the set $\{H,C^{(m)}\}$, $1<m\leq N.$ To be precise, the intermediate Casimir operators $C^{(m)}$ are defined on the 2n-dimensional phase space via 
\be C^{(m)}=\Delta^m(C)\otimes\left(\otimes^{n-m} id\right)\label{Cm}.\ee
The algebra generators $J_i^{(m)}$ can be similarly defined. 
Note that since each successive $C^{(m)}$ includes  a new variable, $x_m$ not appearing in the previous Casimirs, the set will be functionally independent. 

Furthermore, there will exist another set of mutually commuting integrals obtained by embedding the operators $\Delta^m(C)$ in the opposite way, namely defining
\be C_{(m)}=\left(\otimes^{n-m} id\right) \otimes \Delta^m(C),\label{Cmdual}\ee
gives an additional set of commuting integrals. As an example, consider the two-fold copropduct, the resulting operators are 
$$ J_-^{(1)}=x_1^2, \qquad J_{-}^{(2)}=x_1^2+x_2^2, \qquad J_{-}^{(3)}=x_1^2+x_2^2 +x_3^2,$$
$$ J_+^{(1)}=p_1^2, \qquad J_{+}^{(2)}=p_1^2+p_2^2, \qquad J_{+}^{(3)}=p_1^2+p_2^2 +p_3^2,$$
$$ J_3^{(1)}=x_1p_1, \qquad J_{3}^{(2)}=x_1p_1+x_2p_2, \qquad J_{3}^{(3)}=x_1p_1+x_2p_2 +x_3p_3,$$
$$ C^{(1)}=0, \qquad C^{(2)}=(x_1p_2-x_2p_2)^2, \qquad C_{(2)}=(x_3p_2-x_2p_3)^2,$$
and
$$ C^{(3)}=(x_1p_2-x_2p_1)^2+(x_1p_3-x_3p_1)^2+(x_3p_2-x_2p_3)^2.$$
Notice that whereas the $\left(J_{i}\right)_{(m)}$ are linearly dependent on the set $J_{i}^{(m)}$, the Casimirs are not. Indeed, we shall now prove that the set of $2n-3$ functions $\{C^{(m)}, C_{(m)}\}$ are functionally independent and will still be functionally independent when the Hamiltonian is included. 

\begin{theorem}\label{angularintegrals}
For $N> 2$, the set of Casimir operators defined via (\ref{Cm}) and (\ref{Cmdual}) are functionally independent and furthermore the set remains functionally independent when the Hamiltonian is included. 
\end{theorem}
{\bf Proof} The Casimir functions are constructed from linear combinations of the squares of the  $N(N-1)/2$ functionally independent generators of rotations in $ND$. Furthermore, the set of $2N-3$ functions are linearly independent and so the set is also functionally independent.   The inclusion of the Hamiltonian will not change the functional dependence since it depends non-trivially on the radial coordinate $r=\sqrt{J_-^{(n)}},$ while the other $2N-3$ functions depend only on the angular coordinates.

The crucial point is that the coalgebraic structure of a given Hamiltonian is not invariant under a canonical change of variables that intertwines some of the coordinates. This implies that we have the possibility of constructing new coalgebraic systems by applying a change of variable to a 2D quadratically superintegrable system and then extending to ND. For the sake of concreteness let us consider the representation \ref{rep2} in polar coordinates 
\begin{equation*}
 J_-^{(2)} = r^2 , \quad J_3^{(2)} = r p_3 , \quad J_+^{(2)} = p_r^2 + \frac{p_\theta^2}{r^2}.
\end{equation*} 
By changing the winding number of the trajectory (\ref{closedtraj}) using  a canonical change of variable in $\theta$
\begin{equation}
\label{winding}
\theta = \beta \theta', \qquad 
p_\theta = \frac{p_{\theta'}}{\beta}, \qquad  \beta = \frac{\ell}{k} , \quad \ell, k  \in \mathbb{N}
\end{equation}
the new system will close its trajectory after $k$ revolutions.  

\begin{figure}
\centering
\includegraphics [width=0.6\textwidth]{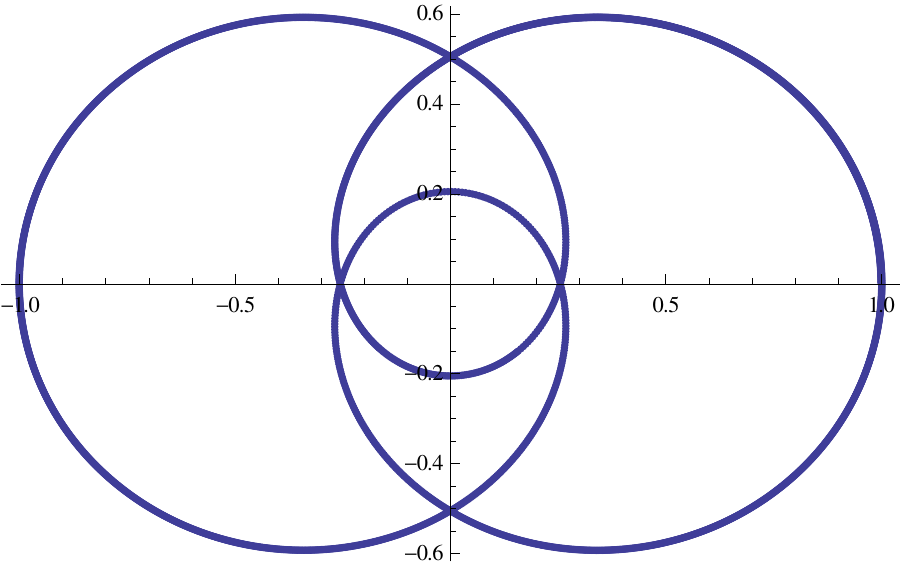}
\caption{$ k=-0.04; \mu = 60 ; \beta= \frac{2}{3}$ }
\label{dueterzi} 
\end{figure}

The change of variable doesn't affect the superintegrability of the system, however it induces a new coalgebraic Hamiltonian, where the Casimir operator has been scaled by a factor of $\beta^{-2}:$ 
\begin{equation}\label{Hbeta}
H_\beta  =  \frac{(1 + k^2 J_-^{(2)})^2}{2} \left( \frac{(J_3^{(2)})^2}{J_-^{(2)}} + \frac{{C}^{(2)}}{ \beta^2 J_-^{(2)}} \right) - \mu \frac{1 - k^2 J_-^{(2)}}{\sqrt{J_-^{(2)}}} 
\end{equation}
Conversely, we can directly observe that multiplying the Casmimir function by a constant will preserve any integrals of motion. However, it may result in a system defined on a new different manifold. 

Indeed these two systems can be connected through a change of variable only in dimension $ N=2$. This becomes evident if the kinetic energy part of the Hamiltonian is considered. In particular, the metric of the original kinetic energy part 
\begin{equation}\label{metric}
\mathcal{T} \rightarrow  ds^2 = \frac{1}{(1 + k^2 J_{-}^{(n)})^2}\left(dr^2 + r^2 d \Omega^2\right) \nonumber,\end{equation}
will be transformed to a new metric 
$$
\mathcal{T}_\beta  \rightarrow ds^2 = \frac{1}{(1 + k^2 J_{-}^{(n)})^2}\left(dr^2 + \beta^2 r^2 d \Omega^2\right),$$
and the corresponding change in the scalar curvature is given by 
$$ R=-4N(N-1) k^2 \rightarrow $$
$$ R_\beta = (N-2)(N-1) (1 - \frac{1}{\beta^2}) \frac{(1 + k^2 r^2)^2}{r^2} - 4N(N-1)k^2,$$
which coincide only if $N=2$.
However both $H$ and $H_\beta$ are still characterized by the closed orbit trajectory in any $N$ since they are algebraically equivalent if projected on the 2D plane which contains the orbit. 
This is a very strong clue about the maximal superintegrability of these classical systems in any dimension $N$. Indeed, in \cite{riglioni2013classical}, higher-order integrals for arbitrary  $\beta$ were constructed  and the MS was explicitly proven in \cite{ballesteros2009hamiltonian} for $N=3.$

In this paper, we consider the quantum case. Namely, instead of a symplectic representation of $\frak{sl}(2)$ we use the canonical quantization to obtain a function space representation 
\be \label{sl2quant} J_{-}=x^2, \qquad J_{+}=-\hbar^2 \partial_x^2, \qquad J_{3}=-i\hbar(x\partial_x+\frac12).\ee
The resulting radial Hamiltonian will be similar to the one given in (\ref{Hbeta}), except with a factor of $\hbar^2$ in front of the kinetic energy term,
\be \fl  \label{HbetaQ} H'=  \frac{(1 + k^2 J_-^{(n)})^2}{2}\left( J_-^{(n)}\right)^{-1}\left( (J_3^{(n)}+i\hbar)^2+ \frac{1}{\beta^2}C^{(n)}\right) - \mu\left(J_-^{(n)}\right)^{-\frac12}\left(1 - k^2 J_-^{(n)}\right).
\end{equation}
We note that it is possible to interpret inverses and roots as Taylor series in the enveloping algebra of the representation for $\frak{sl}(2)$ which will converge away from $r^2=J_-^{(n)}=0.$ 

This quantum system was introduced in  \cite{riglioni2013classical}, where also the superintegrability of (\ref{HbetaQ}) for some specific values of $\beta $ was shown.  However it was not possible to prove the superintegrability for any $\beta$ since  the method of the paper was via direct computations of the commutations relations for each realization of the operators for a fixed $\beta.$

 The main purpose of this paper is to prove the superintegrability of (\ref{HbetaQ}) by providing a coalgebraic analysis at the level of the supersymmetric algebra which characterize the radial system. A similar analysis has also been  introduced in \cite{riglioni2014superintegrablespin} to enlarge the superintegrability properties of some families of superintegrable quantum systems involving spin interaction. In this paper we will show how that formalism provide a natural and straightforward way to prove the MS also for scalar systems. Additionally, we discuss the action of coupling constant metamorphosis for the radial systems and finally, as an example, give yet another proof of the MS of the TTW system via a restriction of the integrals for the radial system. 
 
The paper is organized as follows, in Section 2 we give the extension of the system into ND generated by the coalgebra structure along with the additional, non-radial integrals. The systems are then deformed either by the addition of a winding parameter $\beta$ or via coupling constant metamorphosis. The MS of each system is demonstrated explicitly. In Section 3, we perform dimensional reduction on the 4D system to construct non-radially symmetric systems in 2D which as MS. Section 4 contains concluding remarks.
  
\section{Spectrum generating algebra and superintegrability}
As remarked in the introduction, one of the main properties of superintegrable systems is their exact solvability. In quantum mechanics we can define the exact solvability as the possibility of obtaining, through algebraical methods, the entire spectrum and the set of its eigenfunctions. Such a property is generally exhibited by those Hamiltonian operators which can be solved by factorization method. This method consists in the possibility of factorizing the Hamiltonian  $H = a^\dagger a$ where $a^\dagger$ $a$ act as ladder operators, namely they connect different eigenfunctions. The existence of differential operators mapping eigenfunctions of $H$ into other eigenfunctions of $H$ is indeed typical of superintegrable systems. Let us consider an Hamiltonian $H$ describing an nD quantum system, which admits a set of n constants of the motion in involution $A_i$, $0<i \leq n$
$$
[H,A_i] = [A_i , A_j] = 0
$$
The eigenfunctions of $H$ can be expressed by means of the quantum numbers $a_i$
$$
A_i \psi_{{\mathbf{a_{i,j}}}} = a_i \psi_{{\mathbf{a_{i,j}}}}, \qquad H \psi_{{\mathbf{a_{i,j}}}} = E_j \psi_{{\mathbf{a_{i,j}}}}  
$$
If the system is superintegrable then there exists an extra integral of motion $B$ such that 
\begin{equation}
\label{ei}
[B,H] =0, \qquad [B,A_i] \neq 0. 
\end{equation}
The equations (\ref{ei})  implies that any integral $B$ maps an eigenfunction $\psi_{{\mathbf{a_j}}}$ in a linear combination of isoenergetic  eigenfunctions of $H$
$$
B \psi_{{\mathbf{a_{i,j}}}} = \sum_i c_i \psi_{{\mathbf{a_{i,j}}}}.
$$
Thus, it may be possible to use the ladder operator coming from the spectrum generating algebra of  a given Hamiltonian to define operators which fix the energy eigenspaces of the Hamiltonian and which therefore commute with $H$.    As a concrete example let us consider the $\beta=1$ version of  (\ref{HbetaQ}) which we report here as a function of the angular momentum operator 
\begin{equation}
\label{hl}
H({L}) = \frac{(1+k^2 r^2)^2}{2} \left( p_r^2 - i \frac{\hbar}{r} p_r + \frac{L^2}{r^2} \right) - \mu \left( \frac{1}{r} -k^2 r \right),
\end{equation}
where
$$
p_r = -i \hbar \partial_r, \quad  \quad L = -i \hbar \partial_\phi.
$$
The Hamiltonian (\ref{hl}) exhibits the algebraic structure typical of the shape invariant systems. Namely a factorization type formula
\bea
\label{intertwining}
(L + \frac{\hbar}{2})^2 H(L)  =  A^\dagger  A + {G}, \\
 (L + \frac{\hbar}{2})^2 H(L+\hbar)  =  A A^\dagger + {G}
\eea
which results in $A$ and $A^\dagger$ acting as raising and lowering operators for the angular momentum in the Hamiltonian
\be 
A H(L)  =  H(L+\hbar) A, \qquad 
A^\dagger H(L+\hbar)  =  H(L) A^\dagger.\ee
The operators $A$, $A^\dagger,$ and $  {G}$ are defined as
\bea
\label{ascala}
A^\dagger = \frac{L+\frac{\hbar}{2}}{\sqrt{2}} \left( - (1 + k^2 r^2) p_r + i (L+\hbar)(\frac{1}{r} - k^2 r) \right) - i \frac{\mu}{\sqrt{2}}  \label{AL} \\
A= \frac{L+\frac{\hbar}{2}}{\sqrt{2}} \left( - (1 + k^2 r^2) p_r - i L (\frac{1}{r} - k^2 r) \right) + i \frac{\mu}{\sqrt{2}} \nonumber \\
{G}= 2k^2 L (L + \hbar ) (L + \frac{\hbar}{2} )^2 - \frac{\mu^2}{2}.
\nonumber
\eea
Note that the operators $A$ and $A^\dagger$ are mutual adjoints with respect to the metric (\ref{metric}). 
Since the operators $A^\dagger$ , $A$ act as raising (lowering) operators on the angular momentum $L$, we look for a new couple of operators to balance out this action. The necessary operators are given by 
\begin{equation}
\label{Lpm}
L^+  = e^{i \phi}, \qquad 
L^- = e^{-i \phi}
\end{equation}
which have the required action
\be \label{lpm}
L^+ L = (L - \hbar) L^+, \qquad 
L^- L = (L + \hbar) L^-.\ee
\begin{theorem}
The operators $L^+A$ and $A^\dagger L^{-}$, defined via (\ref{AL}) and (\ref{Lpm}) commute with the Hamiltonian (\ref{hl}). 
\end{theorem}
{\bf Proof}: Taking into account (\ref{intertwining}) and  (\ref{lpm}) it is straightforward to prove  that
$$ L^+A H=L^{+}H(L+\hbar)A=HL^{+}A.$$
A similar computation holds for $A^\dagger L^{-}$ and so, 
\begin{equation*}
[H , L^+ A] = [H , A^\dagger L^-] = 0. 
\end{equation*}
Thus, the operators commute with the Hamiltonian.  

As a consequence, we obtain the following two Hermitian constants of the motion for $H(L)$ 
\begin{eqnarray}\label{R1R2}
\mathcal{R}_+  =   \frac{1}{\sqrt{2}} (L^+ A + A^\dagger L^- ), \qquad 
\mathcal{R}_-  =   \frac{1}{\sqrt{2} i} (L^+ A - A^\dagger L^- ),
\end{eqnarray}
which close to form the following quadratic symmetry algebra 
\begin{equation*}
[ \mathcal{R}_+ , \mathcal{R}_- ]  =  - i \hbar L (2 H -k^2 (8 L^2 + \hbar^2)) 
\end{equation*}
\begin{equation*}
[ \mathcal{R}_+ , L ]  =  - i \hbar \mathcal{R}_-
\end{equation*}
\begin{equation*}
[ \mathcal{R}_- , L ]  =  i \hbar \mathcal{R}_+. 
\end{equation*} 
Note that,  in the flat case $k=0$, the algebra is isomorphic to $\mathfrak{so} (3)$ as the 2-dimensional hydrogen atom.  

\subsection{Coalgebraic extension of the radial systems}
Following the steps outlined in the introduction we can generalize the dimension of the superintegrable system (\ref{hl}) to an arbitrary $N$ by replacing the quantum mechanical representation of $\mathcal{R}_+(-i \hbar \partial_{x_i,} x_i) , \mathcal{R}_-(-i \hbar \partial_{x_i}, x_i), L(-i \hbar \partial_{x_i}, x_i), H(-i \hbar \partial_{x_i}, x_i)$, whose dimension is fixed, 
with functions of coalgebra generators   $J_+^{(N)},J_- ^{(N)},J_3^{(N)}$, thereby constructing a quantum mechanical Hamiltonian in a Darboux space of arbitrary dimension $N$.

Recall that the n-fold coproduct of the representation for $\frak{sl}(2)$ given by  (\ref{sl2quant}) $J_+,J_-,J_3$ 
%
%
by means of which we arrive to the following definition for the Hamiltonian $H$ in $ND$, 
\bea 
H& \equiv & \frac{(1 + k^2 J_-)^2}{2}J_+- \frac{\mu}{\sqrt{J_-}} (1 - k^2 J_-)\nonumber\\
\label{coalgebraicladderb}
 &=&\frac{(1 + k^2 J_-)^2}{2} \left( \frac{1}{J_-} (J_3 +i\hbar)^2 + \frac{L^2}{J_-}\right) - \frac{\mu}{\sqrt{J_-}} (1 - k^2 J_-),
\end{eqnarray}
where the Casimir element is related to $L^2$ via $L^2=C + \hbar^2$ and 
$$ C=\frac12\left(J_+J_-+J_-J_+\right)-J_3^2=J_-J_+-(J_3+i\hbar)^2-\hbar^2.$$
As in the previous equations,  we drop the superscript $(N)$ except where it is important to emphasize the dimension of the underlying space.

We shall now express the operators of the previous section in terms of the coalgebra generators with the ultimate aim of extending into higher dimensions. 
The operators $L^2,$ $H(L),$  $A,$ and $A^\dagger$ can be expressed in terms of the coalgebra generators as 
\begin{eqnarray}
\label{coalgebraicladdera}
\fl L^2 & \equiv &   C+\hbar^2\\
\fl A^\dagger & \equiv & \frac{1}{\sqrt{2}} \left( - (L + \frac{\hbar}{2}) \frac{1 + k^2 J_-}{\sqrt{J_-}} (J_3 + i \hbar) + i \frac{L + \hbar}{\sqrt{J_-}}(L + \frac{\hbar}{2})(1 - k^2 J_-)  \right) - i \frac{\mu}{\sqrt{2}}  \nonumber \\
\fl A & \equiv & \frac{1}{\sqrt{2}} \left( - (L + \frac{\hbar}{2}) \frac{1 + k^2 J_-}{\sqrt{J_-}} (J_3 + i \hbar) - i \frac{L(L + \frac{\hbar}{2})}{\sqrt{J_-}}(1 - k^2 J_-)  \right) + i \frac{\mu}{\sqrt{2}}\nonumber.\eea
Note that when $N=2$, $L^2$ is a perfect square and so the square roots are  defined as simply the generators of rotation, which is formally self-adjoint with respect to the flat metric. However, in higher dimensions, the Casimir operators will not be a perfect square and so this will be addressed in detail later. 

The final operators necessary for the construction of the generalize Runge-Lenz vectors $R_+$ and $R_-$ in $ND$ are the operators $L^+$  and $L^-.$ However the generators of $\mathfrak{sl}(2)$ cannot be used to define these  non-radial operators. In order to overcome this difficulty let us enlarge the coalgebra $\mathfrak{sl}(2)$ as described in the introduction, by including not just the generators for the 2D representation but also the 1D representation. Using these, we can inductively recover the Heisenberg algebra generated by $x_j$ and $p_j$ i.e. $x_j=\sqrt{J_-^{(j)}-J_{-}^{(j-1)}}$ and $p_j= \sqrt{J_+^{(j)}-J_{+}^{(j-1)}}.$ The commutation relations are
 \begin{eqnarray}
\label{algebra}
[ J_3^{(n)} , J_+^{(m)} ] = 2i \hbar J_+^{(\ell)}  \phantom{..} \hspace{1pt} & ; &  [ J_3^{(m)} , J_-^{(n)}] = -2 i \hbar J_-^{(\ell)} \\ 
{ [J_-^{(n)} , J_+^{(m)}  ] } = 4 i \hbar J_3^{(\ell)} \phantom{..} \hspace{1pt} & ; & [ J_3^{(n)} , F(J_\pm^{(m)} ) ] = \pm 2 i \hbar J_\pm^{(\ell)} F'(J_\pm^{(m)}) ,\end{eqnarray}
where $\ell=\min(m,n)$. The commutation relations for the $x_k$ and $p_k$ are identically 0 if $n<k$ and, otherwise, as follows
\bea
\begin{array}{lcr}
{ [ J_+^{(n)} , x_k ]} = -2 i \hbar p_k &  & [J_+^{(n)} , p_k ] = 0 \\
{ [ J_-^{(n)} , x_k ]} = 0 \phantom{a.......} \hspace{2pt} &  & [J_-^{(n)} , p_k ] = 2 i \hbar x_k \\
{ [ J_3^{(n)}, x_k ]} = - i \hbar x_k \hspace{7pt} &  & [J_3^{(n)} , p_k ] = i \hbar p_k \\ \label{algebraend}
{ [ p_k , x_l ]} = -2 i \hbar \delta_{k,l} &  &k\leq n.
\end{array}
\end{eqnarray}

The operator $L^\pm = \frac{x_1}{r} \pm \frac{i x_2}{r}$ can be expressed as a vector if we redefine $L^\pm =  e^{\pm i \phi} \rightarrow e^{\pm i \phi} L. $ In fact, in a three dimensional space $x_2 L$ can be regarded as the first component of the vector product $(0,y,0) \times (0,0,L) = (yL,0,0)$   or in the language of the algebra 
\begin{equation}
x_2 L = x_1 J_3^{(2)} - J_-^{(2)} p_1 + i \hbar x_1.  
\end{equation}
This leads to the coalgebraic definition of the operators $L^\pm$ in 2D as
\begin{eqnarray}
\label{lpmcoalgebraa}
L^- & \equiv & \frac{x_1}{\sqrt{J_-^{(2)}}} L - \frac{i}{\sqrt{J_-^{(2)}}} ( x_1 J_3^{(2)} - J_-^{(2)} p_1 + i \hbar x_1  ), \\
\label{lpmcoalgebrab}
L^+ & \equiv & L \frac{x_1}{\sqrt{J_-^{(2)}}} + \frac{i}{\sqrt{J_-^{(2)}}} ( x_1 J_3^{(2)} - J_-^{(2)} p_1 ). 
\end{eqnarray}
The advantage of this form is that it accommodates an extension to higher dimensions, namely the corresponding operators in $ND$ are chosen as
\begin{eqnarray}
\label{lpmcoalgebraak}
L_k^- & \equiv & \frac{x_k}{\sqrt{J_-^{(n)}}} L - \frac{i}{\sqrt{J_-^{(n)}}} ( x_k J_3^{(n)} - J_-^{(n)} p_k + i \hbar x_k  ), \\
\label{lpmcoalgebrabk}
L_k^+ & \equiv & L \frac{x_k}{\sqrt{J_-^{(n)}}} + \frac{i}{\sqrt{J_-^{(n)}}} ( x_k J_3^{(n)} - J_-^{(n)} p_k ), 
\end{eqnarray}
where $(L_k^+)^\dagger = L_k^-$. Note that in 2D, $L^{\pm}=L_1^{\pm}$ and $L_{2}^+=iL_1^{-}$, $L_2^{-}=-iL_1^{+}.$
\begin{theorem}\label{Lpmt}
The operators $L_k^{\pm}$ defined via (\ref{lpmcoalgebraak}) and (\ref{lpmcoalgebrabk}) satisfy 
$$[L_k^{\pm},L]=\mp \hbar L_k^{\mp}. $$
\end{theorem} 
When $k=1$, $L_{1}^{\pm}$ is exactly the nD extension of the operators $L^{\pm}$. Using the coalgebra extension and the commutation relations (\ref{algebraend}), we see that 
$$\left[L,L^{\pm}\right]=\mp \hbar L^{\pm}\Rightarrow \left[L, L_1^{\pm}\right]=\mp\hbar L_{1}^{\pm}.$$
The identity also holds for other $m$ by permutation symmetry. A direct proof of Theorem \ref{Lpmt} is given in the Appendix.

There does however arise a problem in the definition of $L$ in a number of dimension $N>2$.  According to the definition given in (\ref{coalgebraicladdera}), $L$ is defined as 
\begin{eqnarray}
 L & = & \sqrt{C^{(N)} + \hbar^2} = \sqrt{\sum_{i<j<n} L_{i,j}^2 + \frac{\hbar^2(n-2)}{4}} \\
 && L_{i,j}  \equiv  x_i p_j - x_j p_i.  \nonumber
\end{eqnarray}
Such a problem can be formally solved by defining $L$ as follow
\begin{equation*}
L = - \frac{i}{2} (\gamma_i \gamma_j L_{i,j} + i \hbar (N-2)),\end{equation*}
so that
\be  L^2 = \frac{J_+^{(n)} J_-^{(n)} + J_-^{(n)} J_+^{(n)}}{2} -(J_3^{(n)})^2 + \hbar^2 , 
\end{equation}
where the $\gamma_i$ are anticommuting objects obeying to the algebra
\begin{eqnarray*}
{ [ } \gamma_i , \gamma_j { ] }_+ & = & 2 \delta_{i,j} \\
{[} \gamma_i , J_{+,-,3} {]} & = & {[} \gamma_i , p_i {]} = {[} \gamma_i , x_i {]} = 0   
\end{eqnarray*}
It is with this form of the operators $L$ that the identities for the ladder operators are proven in the Appendix. 

From these concrete realizations of the ladder operators $L_m^{\pm}$ it is possible to construct the ND extension of the Hermitian operators $\mathcal{R}_+$ and $\mathcal{R}_-$ as in (\ref{R1R2}) with $L^{\pm}$ chosen as $L_1^{\pm}$, for example. We therefore have constructed enough functionally independent integrals of the motion to assert that $H$ as defined in (\ref{coalgebraicladderb}) is MS. 
\begin{theorem} The Hamiltonian defined as in (\ref{coalgebraicladderb}) has $2N-1$ algebraically independent integrals of the motion. 
\end{theorem}
As in Theorem \ref{angularintegrals}, the intermediate Casimir elements generated by angular momentum operators $L_{jk}$ will commute with the Hamiltonian  giving $2n-3$ algebraically independent second-order integrals. The additional operators $R_+$ and $R_-$ commute with the Hamiltonian 
\begin{equation}
 [{R}_\pm (\{ J^{(N)} \}) , H( \{ J^{(N)} \})] = [{R}_\pm (\{ J^{(2)} \}) , H(\{ J^{(2)} \}) ] = 0 ,
\end{equation}
for $k=1,2$, and are algebraically independent in dimension $n=2$ and so will be when embedded into ND. Thus, the Hamiltonian has $2N-1$ integrals of motion and is maximally superintegrable.

To obtain scalar integrals, it is possible to simply observe that the operator $L^{+}A$ is linear in the vector $L$ 
\begin{equation}
L^+ A  = \mathcal{A}(\{J\},x_i,p_i) + L \mathcal{B}(\{J\},x_i,p_i)  
\end{equation}  
and commutes with the scalar Hamiltonian 
$$
\nonumber
[H , \mathcal{A}(\{J\},x_i,p_i) + L \mathcal{B}(\{J\},x_i,p_i)   ] = [H ,  \mathcal{A}(\{J\},x_i,p_i)] + L [H, \mathcal{B}(\{J\},x_i,p_i)  ] = 0.
$$
Thus, each of the components of the vector $L^+ A$ must individually commute with the Hamiltonian, i.e. 
\begin{equation}
\label{reductionb}
 [H ,  \mathcal{A}(\{J\},x_i,p_i)] =  [H, \mathcal{B}(\{J\},x_i,p_i)  ]  = 0,
\end{equation}
and analogously for $A^\dagger L^{-}.$ 
Thus, it is possible to build integrals of motion that do not depend on the anticommuting $\gamma_i$. 
In particular, for  $L^+A$, the function $\mathcal{B}(\{J\},x_j,p_j)$ turns out to be the quantum generalization of the Runge-Lenz vector for a Coulomb system on a constant curvature space in agreement with the results previously obtained in \cite{riglioni2013classical} that were obtained without the ladder operator formalism.

\subsection{Canonical transformation and new coalgebraic systems}
As we underlined in the introduction, any point canonical transformation has no effects on the integrability properties of the 2D system. However, the system  in the new variables can induce a new coalgebraic system which is intrinsically a different system when realized in higher dimensions. To be concrete let us apply the angular change of variables (\ref{winding}) to the quantum system  (\ref{hl}) 
$$ H_\beta = \frac{(1 + k^2 r^2)^2}{2} \left( p_r^2 -i\frac{\hbar}{r}p_r + \frac{L^2}{\beta^2 r^2} \right) - \mu \frac{1 - k^2 r^2}{r}, $$
where $ \beta ={m}/{n}$ with $ m,n \in \mathbb{N}.$
The same transformation can be applied to the radial ladder operators $A$ , $A^\dagger$  (\ref{ascala}) to give the more general intertwining relations:
\begin{equation}
\label{intertwiningbeta}
\fl (\frac{L}{\beta} + \frac{\hbar}{2})^2 H_\beta (L)  =  A_\beta^\dagger A_\beta + {G}_\beta(L),\qquad 
 (\frac{L}{\beta} + \frac{\hbar}{2})^2 H_\beta (L+ \hbar \beta)  =  A_\beta A_\beta^\dagger  + {G}_\beta(L),
\end{equation}
which imply
\be
A_\beta H_\beta(L)  =  H_\beta(L+ \hbar \beta) A_\beta,\qquad 
A_\beta^\dagger H_\beta(L+ \hbar \beta )  =  H_\beta (L) A_\beta^\dagger. \label{intertwiningbeta}
\end{equation} 
Here $A_\beta$, $A_\beta^\dagger, {G}_\beta$ are now defined as
\begin{eqnarray*}
 A_\beta^\dagger = \frac{\frac{L}{\beta} + \frac{\hbar}{2}}{\sqrt{2}} \left( - (1 + k^2 r^2) p_r + i (\frac{L}{\beta}+\hbar)(\frac{1}{r} - k^2 r) \right) - i \frac{\mu}{\sqrt{2}},  \\
 A_\beta  = \frac{\frac{L}{\beta} + \frac{\hbar}{2}}{\sqrt{2}} \left( - (1 + k^2 r^2) p_r - i \frac{L}{\beta} (\frac{1}{r} - k^2 r) \right) + i \frac{\mu}{\sqrt{2}}, \\
 {G}_\beta (L) = 2k^2 \frac{L}{\beta} ( \frac{L}{\beta} + \hbar ) ( \frac{L}{\beta} + \frac{\hbar}{2} )^2 - \frac{\mu^2}{2},
\end{eqnarray*}
which induce the following coalgebraic objects
\begin{eqnarray*}
\fl A_\beta^\dagger & \equiv & \frac{1}{\sqrt{2}} \left( - (\frac{L}{\beta} + \frac{\hbar}{2}) \frac{1 + k^2 J_-}{\sqrt{J_-}} (J_3 + i \hbar) + i \frac{\frac{L}{\beta} + \hbar}{\sqrt{J_-}}(\frac{L}{\beta} + \frac{\hbar}{2})(1 - k^2 J_-)  \right) - i \frac{\mu}{\sqrt{2}}, \\
\fl  A_\beta  & \equiv & \frac{1}{\sqrt{2}} \left( - (\frac{L}{\beta} + \frac{\hbar}{2}) \frac{1 + k^2 J_-}{\sqrt{J_-}} (J_3 + i \hbar) - i \frac{ \frac{L}{\beta} (\frac{L}{\beta} + \frac{\hbar}{2})}{\sqrt{J_-}}(1 - k^2 J_-)  \right) + i \frac{\mu}{\sqrt{2}}, \end{eqnarray*}
including the Hamiltonian 
\be \label{Hlbeta} H_\beta   \equiv  \frac{(1 + k^2 J_-)^2}{2} \left( \frac{1}{J_-} (J_3 + i \hbar)^2 + \frac{L^2}{\beta^2 J_-}\right) - \frac{\mu}{\sqrt{J_-}} (1 - k^2 J_-),\ee
already referenced in the introduction (\ref{HbetaQ}). 
The intertwining relations in (\ref{intertwiningbeta}) together with the assumption that  $\beta=m/n$ entail the following ``integer shift" if we consider the power $A^n, (A^\dagger)^n$ instead of $A,A^\dagger$ defined to be 
\be \label{defAm} (A_\beta)^n=A_\beta(L +\hbar(n-1)\beta )\cdots A_\beta(L+ \hbar \beta) A_\beta(L),\ee
\be \label{defAdm} (A^\dagger_\beta)^n=A_\beta^\dagger (L ) \cdots  A_\beta^\dagger(L+ \hbar \beta (n-2) ) A_\beta^\dagger(L+ \hbar \beta (n-1) ).\ee
With this definition, the ladder operators that have the appropriate action on the Hamiltonian functions
\begin{eqnarray*}
(A_\beta  )^n H_\beta(L)  \equiv  H_\beta(L+\hbar m) (A_\beta)^n, \\
(A_\beta^\dagger)^n H_\beta(L+\hbar m)  \equiv  H_\beta (L) (A_\beta^\dagger )^n,
\end{eqnarray*}
which hold according to the ladder relations(\ref{intertwiningbeta}). 
Analogous to what we have seen in the previous sections, it is possible to balance the action of $A^n, (A^\dagger)^n$ by using the counteraction of $L_k,L_k^+$ (\ref{lpmcoalgebraa}, \ref{lpmcoalgebrab}). So we can define the new constants of the motion for $H_\beta(L)$ as follows:
\begin{eqnarray}
\label{betarunge}
{R}_{\beta,1} = \frac{1}{ \sqrt{2}} \left( (L_k^+)^m (A_\beta) ^n +(A_\beta^\dagger)^n (L_k^-)^m \right),\\
{R}_{\beta,2} = \frac{1}{i \sqrt{2}} \left( (L_k^+)^m (A_\beta) ^n - (A_\beta^\dagger)^n (L_k^-)^m \right). 
\end{eqnarray}
It should be stressed that unless we are in the case of dimension 2, where $L^{\pm}$ admit a zeroth-order representation as a differential operator, the objects (\ref{betarunge}) are higher order constants of the motion of order $m+2n$ which can be reduced to the order $m+2n-1$ as described in (\ref{reductionb}). These results hold for all values of $\beta \in \mathbb{Q}$ and define a infinite class of superintegrable systems with higher order constants of the motion in spaces of arbitrary dimension $N$.  
 
\subsection{Spectrum generating algebra for superintegrable systems obtained through the application coupling constant metamorphosis }

The Coupling Constant Metamorphosis (CCM) \cite{HGDR, KMP10, Post20111} is a transformation which puts in correspondence two classes of superintegrable systems. In particular such a transformation can be used as an algorithm to generate the constants of the motion of one superintegrable system given the symmetries of its CCM partner. It is known that all the superintegrable systems on the 2-dimensional Darboux spaces (which are the only ones with a nonconstant scalar curvature admitting second order constants of the motion) can be generated through a CCM transformation applied to a system defined on a space of constant curvature \cite{KKM20042}. The main goal of this section is to show how the structure of the spectrum generating algebra given by the ladder operators $A$ $A^\dagger$ is preserved by the application of the CCM transformation. CCM can be briefly described as an algebraic transformation whose net effect is to exchange the role of the energy and the coupling constant of the system,
\begin{eqnarray}
\label{hccm}
{H} \psi = (\mathcal{T} - \mu \mathcal{V} ) \psi \\
\hat{{H}} \psi \equiv \mu \psi = \frac{1}{\mathcal{V}} (\mathcal{T} - \mathcal{H} ) \psi
\end{eqnarray}
Such a transformation applied to the system (\ref{coalgebraicladderb}) returns the following new Hamiltonian 
\begin{equation}\label{coalgebraicladderccm}
\fl \hat{H}= \frac{\sqrt{J_-}(1 + k^2 J_-)^2 }{2(1 - k^2 J_- -4 \delta \sqrt{J_-} )} (\frac{1}{J_-}(J_3 + i\hbar)^2 + \frac{L^2}{\beta^2 J_-}) + \frac{\tilde{\mu} \sqrt{J_-}  }{2 (1 - k^2 J_- -4 \delta \sqrt{J_-} )}
\end{equation}
where the transformation has been done using as potential $\mathcal{V} = \frac{1}{r} - k^2 r -4 \delta$. It is straightforward to see from ( \ref{hccm} ) which the two Hamiltonians ${H} , \hat{H}$ share the same eigenfunctions except for the exchange ${H} \leftrightarrow \mu $. This entails that the ladder operators defined for the eigenfunctions of ${H}$ work also for the eigenfunctions of $\hat{H}$ provided that $\mu \rightarrow \hat{H} $ is well defined as a differential operator. So we arrive to the following definition for $\hat{A}$ $\hat{A}^\dagger$ which we will provide directly in terms of coalgebraic elements:

\begin{eqnarray*}
\fl \hat{A}_{\beta}^\dagger  &\equiv&  \frac{1}{\sqrt{2}} \left( \! - (\frac{L}{\beta} + \frac{\hbar}{2}) \frac{1 + k^2 J_-}{\sqrt{J_-}} (J_3 + i \hbar) + i \frac{\frac{L}{\beta} + \hbar}{\sqrt{J_-}}(\frac{L}{\beta} + \frac{\hbar}{2})(1 - k^2 J_-)  \! \right) \! - \! i \frac{\hat{H}(L + \hbar \beta)}{\sqrt{2}} \\
\fl \hat{A}_{\beta}  &\equiv&  \frac{1}{\sqrt{2}} \left( \! - (\frac{L}{\beta} + \frac{\hbar}{2}) \frac{1 + k^2 J_-}{\sqrt{J_-}} (J_3 + i \hbar) - i \frac{ \frac{L}{\beta} (\frac{L}{\beta} + \frac{\hbar}{2})}{\sqrt{J_-}}(1 - k^2 J_-)  \! \right) \! + \! i \frac{\hat{H}(L)}{\sqrt{2}}
\end{eqnarray*}  
fulfilling the intertwining relations 
\begin{eqnarray*}
\hat{A}_{ \beta}   \hat{H}_{ \beta} (L)  =  \hat{H}_{ \beta} (L+\hbar \beta ) \hat{A}_{ \beta} \\
\hat{A}_{ \beta}^\dagger \hat{H}_{\beta} (L+\hbar \beta)  =  \hat{H}_{ \beta} (L) \hat{A}_{ \beta}^\dagger 
\end{eqnarray*}
which, analogously to (\ref{betarunge}), determine the following set of vectors for $\hat{H}$ for any $k=1\ldots N$
\begin{eqnarray*}
\hat{R}_{\beta,+} = \frac{1}{ \sqrt{2}} \left( (L_k^+)^m (\hat{A}_ {\beta}) ^n + (\hat{A}_{ \beta}^\dagger)^n (L_k^-)^m \right) ,\\
\hat{R}_{\beta,-} = \frac{1}{i \sqrt{2}} \left( (L_k^+)^m (\hat{A}_{ \beta}) ^n - (\hat{A}_{ \beta}^\dagger)^n (L_k^-)^m \right). 
\end{eqnarray*}
Here again, the m-fold product of the ladder operators requires a shift in $L$ after each application and is defined as in (\ref{defAdm}) and (\ref{defAm}). 
\section{Dimensional reduction and curved superintegrable extensions of the TTW/PW systems}
As an example, we consider the  algebraic Hamiltonians (\ref{coalgebraicladderb},\ref{coalgebraicladderccm}) in 4d.   In this case the two Hamiltonians depend on the representation of $sl_2$ in 4D, i.e. on the operators
\begin{eqnarray}
J_+^{(4)} = -\hbar^2 ( \partial_1^2 + \partial_2^2 + \partial_3^2 + \partial_4^2 ) \\ 
J_-^{(4)} = x_1^2 + x_2^2 + x_3^2 + x_4^2 \\
J_3^{(4)} = -i \hbar (x_1 \partial_1 + x_2 \partial_2  + x_3 \partial_3 + x_4 \partial_4 ) -2i \hbar. 
\end{eqnarray}
It is possible to reduce this representation to one in two variables, as in \cite{RTW2008, RTW2009}, by using a bi-polar coordinates system:
\begin{eqnarray}
\label{variabiliridotte}
x_1 = r_1 \cos \phi_1 & \quad ; \quad x_2 = r_1 \sin \phi_1 \\ \nonumber
x_3 = r_2 \cos \phi_2 & \quad ; \quad x_4 = r_2 \sin \phi_2 , \nonumber
\end{eqnarray}
so that the operators become
\begin{eqnarray}
{J}_-^{(4)} & = &r_1^2 + r_2^2, \qquad 
{J}_3^{(4)} = -i \hbar (r_1 \partial_{r_1} + r_2 \partial_{r_2} + 2) ,\nonumber \\
{J}_+^{(4)} & =& -\hbar^2 (\partial_{r_1}^2 + \frac{1}{r_1} \partial_{r_1} + \frac{1}{r_1^2} \partial_{\phi_1}^2  + \partial_{r_2}^2 + \frac{1}{r_2} \partial_{r_2} +\frac{1}{r_2^2} \partial_{\phi_2}^2 ).\nonumber
\end{eqnarray}

Since the generators are independent on the new angular variables $\phi_1$ , $\phi_2$, it is possible 
to get rid of the two degrees of freedom coming from $\phi_1,\phi_2$ and to obtain a new two dimensional system which inherits the properties of the original 4-dimensional one.
Let us show in a few algebraic steps how to perform the reduction for the quantum case. 
\begin{equation}
\label{version4d}\fl 
< \psi(x_1,x_2,x_3,x_4) | \hat{H}_{4d} | \psi (x_1,x_2,x_3,x_4) > =
 \int  f r_1 r_2 \psi^* \hat{H}_{4d} \psi dr_1 dr_2 d \phi_1 d \phi_2.\end{equation}
Separating the wave function as $ \sqrt{r_1 r_2} \psi (r_1,r_2,\phi_1,\phi_2) = \tilde{\psi} (r_1, r_2) e^{i l_1 \phi_1} e^{i l_2 \phi_2} $ transforms the form of the Hamiltonian (\ref{version4d}) to
\begin{equation}
\int f \tilde{\psi} (r_1, r_2)^* \hat{\tilde{H}}_{2d} \tilde{\psi} (r_1, r_2) dr_1 dr_2,  
\end{equation}
where the reduced (self-adjoint) operator is defined by:
\begin{equation}
\label{reduction}
\fl \hat{\tilde{H}}_{2d} = \sqrt{r_1 r_2}\left( \frac{1}{4 \pi^2} \int e^{-i l_1 \phi_1} e^{-i l_2 \phi_2} \hat{H}_{4d} e^{i l_1 \phi_1} e^{i l_2 \phi_2}  d\phi_1 d \phi_2 \right) \frac{1}{\sqrt{r_1 r_2}}.
\end{equation}
The new Hamiltonian can be now written in terms of the reduced form of the generators $${J}_i^{(4)}\rightarrow {\tilde{J}}_i^{(4)} = \sqrt{r_1 r_2}(\frac{1}{4 \pi^2} \int e^{-i l_1 \phi_1} e^{-i l_2 \phi_2} \hat{J}_i^{(4)} e^{i l_1 \phi_1} e^{i l_2 \phi_2} d \phi_1 d \phi_2 ) \frac{1}{\sqrt{r_1 r_2}}$$ 
which become
\bea
{\tilde{J}}_-^{(4)} & =& r_1^2 + r_2^2 = {J}_-^{(2)}, \label{nonradialqr} \\
{\tilde{J}}_3^{(4)} &= &-i \hbar (r_1 \partial_{r_1} + r_2 \partial_{r_2} + 1) = {J}_3^{(2)}, \\
{\tilde{J}}_+^{(4)} & =& -\hbar^2 (\partial_{r_1}^2 + \frac{1-4 l_1^2}{4 r_1^2}   + \partial_{r_2}^2  + \frac{1-4l_2^2}{4 r_2^2} )\nonumber\\
& =& {J}_+^{(2)} + \hbar^2 \frac{b_1^2}{r_1^2} + \hbar^2 \frac{b_2}{r_2^2},\qquad 
\label{endnonradialqr}
b_1 = l_1^2 - \frac{1}{4}, b_2= l_2^2 - \frac{1}{4}.
\eea

After the reduction, the 4-dimensional representation coincides with the 2-dimensional ones except for the generator $\hat{J}_+$ wich has an additional centrifugal term depending on the quantum numbers $l_1, l_2$ coming from the degrees of freedom we had cut off previously.
The representation (\ref{nonradialqr} - \ref{endnonradialqr}) can also be regarded as a non-radial generalization of  (\ref{rep2}). In particular the representation  (\ref{nonradialqr} - \ref{endnonradialqr}), together with the change of variables 
\begin{eqnarray*}
r_1 = r \cos \frac{\theta}{\beta},\qquad 
r_2 = r \sin \frac{\theta}{\beta}, \nonumber
\end{eqnarray*} 
turns the system (\ref{Hlbeta}) into 
\begin{eqnarray}
\label{pw}
 H_\beta & = & \frac{(1 + k^2 r^2)^2}{2} \left( \Delta_{\mathbb{R}^2} + \frac{b_1^2}{r^2 \cos^2 \frac{\theta}{\beta}} +\frac{b_2^2}{r^2 \sin^2 \frac{\theta}{\beta}} \right) - \mu \frac{1 - k^2 r^2}{r}. 
\end{eqnarray}
This system can be interpreted as a generalization (identical for $k=0$) of the PW system \cite{PW2010} on a space of constant curvature, which has been obtained as a reduction of a system on a space of non-constant curvature $\beta \neq 1$.

Along the same direction it is straightforward to obtain a generalization of the TTW system to a non-flat Riemannian space of Darboux type. Let us consider the system (\ref{coalgebraicladderccm}) and the following change of variables (see \cite{PW2010} for CCM in polar coordinates)
\begin{eqnarray*}
r_1 = r^2 \cos \frac{2 \theta}{\beta},\qquad 
r_2 = r^2 \sin \frac{2 \theta}{\beta}.
\end{eqnarray*} 
The resulting Hamiltonian is an extension of the TTW system (identical at $k=0$) given by the following Hamiltonian
\begin{equation}
\label{ttw}
\fl \hat{H}_\beta = \frac{(1 + k^2 r^4)^2 }{8 (1 - k^2 r^4 -4 \delta r^2 )} \left( -\hbar^2 \Delta_{\mathbb{R}^2}+ \frac{b_1^2}{r^2 \cos^2 \frac{2 \theta}{\beta}} +\frac{b_2^2}{r^2 \sin^2 \frac{2 \theta}{\beta}} \right) + \frac{\tilde{\mu} r^2  }{2 (1 - k^2 r^4 -4 \delta r^2 )}.
\end{equation} 
It is now well established that both of these systems are superintegrable when $k=0$ for rational values of $\beta$. Clearly, both of these systems are integrable, associated with separation of variables in polar coordinates.  In the following section, we shall see how the integrals for the system in 4D can be reduced to give an addition integral for the 2D system, thus proving that both systems (\ref{pw}) and (\ref{ttw}) are MS. Thus we obtain an additional proof of the superintegrability of the TTW system via dimensional reduction.

\subsection{Integrals of the motion for the reduced systems }
As the Hamiltonian itself is reduced, so too can its integrals of the motion can be obtained by a proper reduction of the integrals belonging to the 4-dimensional versions (\ref{coalgebraicladderb} - \ref{coalgebraicladderccm}).
As we showed in (\ref{nonradialqr} - \ref{endnonradialqr}) the elements $J_-, J_3, J_+$ can be easily reduced to 2-dimensional differential operators, however in order to obtain also the integrals of the motion we need to reduce also the  ladder operators $L^+_i$ , $L^-_i$ and $L$.
The differential operators $L^+_j$ $L_j^-$ are linear in $x_j$, $p_j$  therefore their square turns out to be linear in the coalgebraic elements $\{ x_j^2, p_j^2, [x_j, p_k] \}$ and hence  the linear combination $(L_1^\pm)^2 + (L_2^\pm)^2 $ , $(L_3^\pm)^2 + (L_4^\pm)^2 $ are respectively functions of 
\begin{eqnarray}
\label{var1}
\{ J^{(4)}, J^{(2)},  (x_1 \gamma_1 + x_2 \gamma_2),(p_1 \gamma_1 + p_2 \gamma_2) \}\\
\label{var2}
\{ J^{(4)}, J_{(2)},  (x_3 \gamma_3 + x_4 \gamma_4) , (p_3 \gamma_3 + p_4 \gamma_4).\}
\end{eqnarray} 

The first two sets of elements can be reduced by means of the change of variable (\ref{variabiliridotte}) whose application makes the elements (\ref{var1}), (\ref{var2}) independent on the variables $\phi_1$ $\phi_2,$ however the elements depending in the matrices $\gamma$ cannot be so easily reduced. However we can finalize the reduction by introducing a proper similarity transformation.
To be concrete let us introduce the following basis for the $\gamma$ matrices 
\begin{eqnarray*}
\gamma_1 \equiv \sigma_1 \otimes 1 \!\! 1 , \qquad  \gamma_3 \equiv \sigma_3 \otimes \sigma_1 \\
\gamma_2 \equiv \sigma_2 \otimes 1 \!\! 1 , \qquad  \gamma_4 \equiv \sigma_3 \otimes \sigma_2
\end{eqnarray*}
and let us also define the following the following gauge matrix 
\begin{equation*}
R = \left( \begin{array}{cccc}
e^{-\frac{1}{2}(\phi_1 + \phi_2)} & 0 & 0 & 0 \\ 0 & e^{-\frac{1}{2}(\phi_1 - \phi_2)}  & 0 & 0 \\ 0 & 0 & e^{\frac{1}{2}(\phi_1 - \phi_2)}  & 0 \\
0 & 0 & 0 & e^{\frac{1}{2}(\phi_1 - \phi_2)}  
\end{array} \right). 
\end{equation*}
The above gauge turns the reduction (\ref{reduction}) into 
\be \fl \label{reducedrep}
I \rightarrow \tilde{I} = \sqrt{r_1 r_2}(\frac{1}{4 \pi^2} \int e^{-i l_1 \phi_1} e^{-i l_2 \phi_2} R^{-1} I R e^{i l_1 \phi_1} e^{i l_2 \phi_2} d \phi_1 d \phi_2 ) \frac{1}{\sqrt{r_1 r_2}},\ee
by means of which we obtain the following reduced operators. The terms that are linear in the gamma matrices reduce to
\begin{eqnarray*}
x_1 \gamma_1 + x_2 \gamma_2 \rightarrow r_1 \gamma_1, \qquad  &x_3 \gamma_3 + x_4 \gamma_4 \rightarrow r_2\gamma_3  \\
\gamma_1 p_1 \! + \! \gamma_2 p_2 \rightarrow \gamma_1 pr_1 +\frac{\hbar l_1}{r_1} \gamma_2, \qquad &\gamma_3 p_3 \! + \! \gamma_4 p_4 \rightarrow \gamma_3 pr_2 \! + \!  \frac{\hbar l_2}{r_2} \gamma_4,\end{eqnarray*}
and the coalgebraic generators become
\begin{eqnarray*}\nonumber
  J_- \rightarrow r_1^2 + r_2^2 \\ \nonumber
J_3 \rightarrow r_1 pr_1 + r_2 pr_2 -i \hbar \\ \nonumber
 J_+ \rightarrow pr_1^2 \! + \! pr_2^2 \! + \! \frac{\hbar^2 l_1^2}{r_1^2} \! + \! \frac{\hbar^2 l_2^2}{r_2^2} \! + \! i \frac{\hbar^2 l_1}{r_1^2} \gamma_1 \gamma_2 \!  + \! i  \frac{\hbar^2 l_2}{r_2^2} \gamma_3 \gamma_4   \\L \! \rightarrow \! -i \gamma_1 \gamma_3 (r_1 pr_2 \! - \! r_2 pr_1 ) \! -i \hbar \left( l_1 \gamma_1 \gamma_2 \! + \! l_2 \gamma_3 \gamma_4 \! - \! \frac{l_1 r_2}{r_1} \gamma_2 \gamma_3 \! + \! \frac{r_1 l_2}{r_2} \gamma_1 \gamma_4 \right). \nonumber
\end{eqnarray*}
Let us emphasize that the action of this reduction has turned the scalar operator $J_+$ into the following diagonal operator 
\begin{equation*}
\left(
\begin{array}{cccc}
J_+(l_1-1,l_2-1) & 0 & 0 & 0 \\
0 & J_+(l_1-1,l_2) & 0 & 0 \\
0 & 0 & J_+(l_1 , l_2-1) & 0 \\
0 & 0 & 0 & J_+(l_1,l_2)
\end{array} \right),  
\end{equation*}
where $J_+(l_1,l_2)$ is defined as follows
\begin{equation*}
J_+(l_1,l_2) \equiv  pr_1^2 \! + \! pr_2^2 \! + \! \frac{\hbar^2 l_1 (l_1 +1)}{r_1^2} \! + \! \frac{\hbar^2 l_2 (l_2 + 1)}{r_2^2}.
\end{equation*}

As a consequence the Hamiltonian operators (\ref{coalgebraicladderb})  (\ref{coalgebraicladderccm}) turn into the following diagonal ones 
\begin{equation*}
\fl H_{i,j} = \hbar^2 F(r) \left( - \Delta_{\mathbb{R}^2} + \frac{l_1(l_1 + \delta_{i,j} \epsilon^{(1)}_j)}{\beta^2 r^2 \sin^2 \beta\theta} + \frac{l_2 (l_2 + \delta_{i,j} \epsilon^{(2)}_j)}{\beta^2 r^2 \cos^2\beta \theta} \right) + V(r)
\end{equation*}
where
\begin{eqnarray*}
\epsilon_j^{(1)} = (-1,-1,1,1) & ; & \epsilon_{j}^{(2)} = (-1,1,-1,1) \\
r_1 = r \cos \beta \theta & ; & r_2 = r \sin\beta \theta
\end{eqnarray*}
and 
\[F(r) = \frac{(1 + k^2 r^2)^2}{2}, \qquad  V(r)= \frac{- \mu (1 - k^2 r^2)}{r},\]
for the system (\ref{coalgebraicladderb}) and 
\[ F(r) = \frac{r(1 + k^2 r^2)^2}{2 (1 -k^2 r^2 -4 \delta r)} , \qquad V(r) = \frac{-\mu r}{(1 - k^2 r^2 - 4 \delta r)}\]
for the system (\ref{coalgebraicladderb}).

From the above analysis we conclude that it is possible to reduce all the elements of the spectrum generating algebra, but we have to pay as a price that the reduced ladder operators acting on $L$  
can shift $L$ by steps of $2 \hbar$ instead of $\hbar$ since their reduced version is given by quadratic combination of the original operators, defined as 
$$ L^\pm_{r,1}=(L_1^{\pm})^2+(L_2^{\pm})^2, \qquad L^\pm_{r,2}=(L_3^{\pm})^2+(L_4^{\pm})^2.$$
The integrals of the motion for the reduced version of (\ref{coalgebraicladderb}) (\ref{coalgebraicladderccm}), under the representation (\ref{reducedrep}), are constructed via 
\begin{equation*}
\mathcal{R}_{\beta,+} = \frac{1}{\sqrt{2}} \left( (L_{r,j}^+)^{\frac{m'}{2}} (A_\beta)^{n'} +(A_\beta^\dagger)^{n'}(L_{r,j}^-)^{\frac{m'}{2}} , \right) 
\end{equation*}
\begin{equation*}
\mathcal{R}_{\beta,-} = \frac{1}{i\sqrt{2}} \left( (L_{r,j}^+)^{\frac{m'}{2}} (A_\beta)^{n'} -(A_\beta^\dagger)^{n'}(L_{r,j}^-)^{\frac{m'}{2}} , \right) 
\end{equation*}
where $\beta = \frac{m}{n}$ and $ m' = m, n' = n$ for even $m$ and $ m' = 2m, n' = 2n $ for odd $m$. The constant of the motion for the extension of the TTW system, can be constructed analogously with $\hat{A}_m$ and $\hat{A}^\dagger_m.$ Recall, that repeated application of the ladder operators $ A$ and $A^\dagger$ are defined via (\ref{defAdm}) and (\ref{defAm}).

\section{Conclusions}
In this paper, we have shown that the coalgebra formalism is an effective method for extending Hamiltonians into higher dimensions while preserving integrals of the motion, even those integrals that are not expressible entirely in terms of the co-algebraic generators. 
In particular, we discuss the quantum extension of the Coulomb and oscillator type systems on the N-dimensional extension of the so called "Bertrand spaces", in their conformally flat form as introduced in \cite{Bertspacetime}. We prove that such extensions are maximally superintegrable by constructing directly $2N-3$ intermediate Casimir operators from the coalgebra generators as well as two additional higher-order integrals of motion from ladder operators. Thus, we have proven the superintgrability of the systems proposed in \cite{riglioni2013classical}. 

As an example, we have also considered the system in 4D and its dimensional reduction. In order to construct integrals of motion that reduce appropriately, we have introduced a non-trivial gauge leading to a vector of partner Hamiltonians with vector integrals of the motion. As a consequence, we have given on the one hand an alternative proof of the superintegrability of the TTW, on the other hand  a generalization of TTW on Darboux spaces. For earlier proofs, see also \cite{CQ10, CG, KMPog10, KKM10a}. 

We conclude by highlighting a few of the more unique methods employed to construct the integrals. As remarked above, the co-algebra structure was used to extend the system, as well as many of the integrals, to higher-dimensions. However, even the non-radial operators such as $L_k^\pm$ were able to be extended by casting them in appropriate form as in (\ref{lpmcoalgebraak}). Additionally, we have used (formally) a vector form of the angular momentum operator $L$ in order to express the Hamiltonian in factorized form. Usually, the factorization method is applied only for 1D Hamiltonians and extended to higher-dimensions by separation of variables, see e.g. \cite{marquette2009super, Marquette2012, Celeghini2013}. In this paper, the factorization method is applied to the entire 2D and, by extension, ND Hamiltonian.

 \section{Acknowledgments}
D.R. acknowledges a fellowship from the laboratory of Mathematical Physics of the Centre de Recherches Math\'ematiques CRM. S.P. acknowledges funding from the College of Natural Sciences at UH and thanks the CRM for their hospitality during collaborative visits.

\section*{Appendix}
In this appendix, we prove the asserted forms of the ladder operators for total angular momentum in dimension $n$. 

We define the total angular momentum operator, formally, as a vector via 
\be L=\frac{\hbar (n-2)}{2}-\frac{i}{2}\sum_{j=1}^n\sum_{k=1}^n \gamma_j\gamma_k L_{jk}.\ee
\begin{theorem}
The square of the angular momentum vector is the scalar total angular momentum. 
\end{theorem}
\noindent {\bf Proof:} The square of the proposed vector is 
\be\fl  \label{Lsq} L^2=\frac{\hbar^2(n-2)}{4}-\frac{i\hbar (n-2)}{2}\sum_{j=1}^n\sum_{k=1}^n \gamma_j\gamma_k L_{jk} -\frac14\sum_{j,k, \ell, m} \gamma_j\gamma_k \gamma_{\ell} \gamma_{m} L_{jk}L_{m,n}.\ee
The third term of the right-hand side of (\ref{Lsq}) can be decomposed into three cases, all 4 indices are distinct, exactly two  coincide, or two pairs coincide. Recall that $j=k$ and $m=n$ terms are 0, i.e.   $L_{jj}=0$ by definition. In the first case, we have 
$$ \sum_{j\ne k\ne  \ell\ne m} \gamma_j\gamma_k \gamma_{\ell} \gamma_{m} L_{jk}L_{m,n}=\sum_{a< b< c<d} \gamma_a\gamma_b\gamma_c\gamma_d\left(4L_{ab}L_{cd}-4L_{ac}L_{bd}+4L_{ad}L_{bc}\right).$$
However, as can be directly verified from the definition of the $L_{jk}$, the sum is zero for all $a<b<c<d$, i.e. 
\be \label{Ljksym}  L_{ab}L_{cd}-L_{ac}L_{bd}+L_{ad}L_{bc}=0, \qquad a\ne b\ne c \ne d.\ee
For the second case, namely when $j=m$ or $\ell$ and $k=m$ or $\ell$, the sum restricts to
$$ -\frac12 \sum_{j,k, m}  \gamma_k  \gamma_{m}  L_{kj}L_{jm}+\gamma_j \gamma_{m}L_{jk}L_{km}=i\hbar\sum_{k,m}\gamma_k\gamma_mL_{km}\sum_{j\ne k\ne m}1.$$
For the second identity, the commutator of $L_{jk}$ and $L_{km}$ is used 
$$ [L_{jk},L_{km}]=-i\hbar L_{jm}.$$
These terms will exactly cancel the linear terms in (\ref{Lsq}). 
The last non-zero cases are those where $j=\ell, k=m$ or $j=m, k=\ell$ which give 
$$ -\frac14\sum_{j,k} \gamma_j\gamma_k\gamma_j\gamma_k L_{jk}^2+\gamma_j\gamma_k\gamma_k\gamma_j L_{jk}L_{kj}=\sum_{j<k} L_{ij}^2.$$
Thus, the square of $L$ as in (\ref{Lsq}) reduces to 
\be L^2=\frac{\hbar^2(n-2)}{4}+\sum_{j<k} L_{ij}^2.\ee

Next, we construct the ladder operators for $L$, $L^{+}$ and $L^{-}. $ Along the way, we collect some facts. 

\begin{lemma}
The multidimensional analog of $x_2L=x_1J_3^{(2)}-J_-^{(2)}p_1+\frac{i}{2}x_1$ holds. Namely, 
\be \label{sumintoxi} \sum_{j=1}^n {x_j}L_{mj}=x_mJ_3^{(n)}-J_-^{(n)}p_m+\frac{i\hbar nx_m}{2}\equiv R_m.\ee
\end{lemma}
\noindent {\bf Proof:} A direct calculation shows that the left hand-side of (\ref{sumintoxi}) is 
$$ x_mJ_3^{(n)}-J_-^{(n)}p_m+\frac{i\hbar n}{2}x_m=-i\hbar x_m\left(\sum_{j=0}^n x_j\partial_j +\frac{n}{2}\right)+i\hbar \sum_{j=1}^n x_j^2\partial_m +\frac{i\hbar nx_m}{2}.$$
After simplifying this expression, we see that the right-hand-side is simply $\sum_{j=0}^n {x_j}L_{mj}$ and so the lemma is proved. 

\begin{lemma} \label{xmLLemma}
The following identity holds
\be \label{xkL} [x_mL,L]=\frac{\hbar(n-2)}{2}[x_m,L]-i\hbar R_m-i\sum_{j\ne m} x_j[L_{mj}, L].\ee
\end{lemma}
\noindent {\bf Proof:} First, we compute 
\be [x_m, L]=\frac{-\hbar}{2}\sum_{j=1}^n (\gamma_m\gamma_j-\gamma_j\gamma_m) x_j.\ee
Thus, left-hand side of (\ref{xkL}) becomes
\bea [x_m L, L]&=&\frac{-\hbar}{2}\sum_{j=1}^n(\gamma_m\gamma_j-\gamma_j\gamma_m)x_j\left(\frac{\hbar (n-2)}{2}-\frac{i}{2}\sum_{k=1}^n\sum_{\ell=1}^n \gamma_k\gamma_\ell L_{k\ell}\right)\nonumber\\
&=&\frac{\hbar(n-2)}{2}[x_m,L]+\frac{i\hbar}{2}\gamma_m\sum_{j\ne m,k, \ell}\gamma_j\gamma_k\gamma_\ell x_jL_{k\ell} \nonumber .\eea
As above, we separate the sum into cases. The first case is $j\ne k \ne \ell\ne m.$ Ordering the indices sequentially $a<b<c$ gives 
$$ \sum_{j\ne k\ne \ell}\gamma_j\gamma_k\gamma_\ell x_jL_{k\ell}=\sum_{a<b<c}\gamma_a\gamma_b\gamma_c\left(2x_aL_{bc}-2x_bL_{ac}+2x_cL_{ab}\right). $$
Again, by direct computation, we see that the sums inside the parenthesize are identically 0, 
\be \label{xLs} x_aL_{bc}-x_bL_{ac}+x_cL_{ab}=0,\ee
and so the terms of the original sum with all distinct indices are 0. Of course, the cases where $k=\ell$ are identically 0 from the definition of $L_{k\ell}$ and so the sum reduces to the terms with $k=j$ or $\ell=j$ and with $k=m$ or $\ell=m$, which give 
$$\sum_{j\ne m,k, \ell}\gamma_m\gamma_j\gamma_k\gamma_\ell x_jL_{k\ell}=2\sum_{j\ne m}\sum_{k=1}^n\left(-\gamma_m\gamma_k x_jL_{kj}+\gamma_j\gamma_kx_jL_{km}\right).$$
The terms of this sum with $k=m$ are exactly the right hand side of (\ref{sumintoxi}) and so can be expressed in terms of the radial operators and $x_m$, i.e. as the component $R_m$.  The remaining terms can be recognized as follows
\bea \sum_{j\ne m} x_j[L_{mj}, L]=\hbar\sum_{k\ne m,j\ne m}\gamma_k\gamma_m x_jL_{jk}+{\hbar}\sum_{j\ne m, k\ne j}\gamma_j\gamma_k x_j L_{mk}.\nonumber\eea
 So, finally  the commutator is computed as 
$$[x_m L, L]=\frac{-\hbar^2(n-2)}{4}[x_m,L]-i\hbar R_m-i\sum_{j\ne m} x_j[L_{mj}, L].$$
\begin{lemma} \label{RmLlemma}
The following identity holds
\be [R_m, L]=i\hbar x_m L-\frac{i\hbar^2(n-2)}{2}x_m+\sum_{j=1}^nx_j[L_{mj}, L].\label{RmL}\ee
\end{lemma}
\noindent{\bf Proof:} The computation is 
$$ [\sum_{j=1}^n {x_j}L_{mj}, L]=\sum_{j=1}^n\left([x_j, L]L_{mj}+x_j[L_{mj}, L]\right).$$
Beginning with the first terms, gives 
\bea\fl  \sum_{j=1}^n \left[x_j, L\right]L_{mj}&=&\sum_{j}\sum_{k\ne j} \frac{-\hbar}{2}\gamma_{j}\gamma_k x_kL_{mj}\nonumber\\
\fl &=&-\hbar\left( \sum_{j} \frac{1}{2} x_m L_{mj}\gamma_m\gamma_j+\sum_{a<b}\gamma_a\gamma_b\left( x_aL_{mb}-x_bL_{ma}\right)\right).\nonumber
\eea 
The terms in the final parenthesize can be simplified via (\ref{xLs}) leading to 
\bea \sum_{j=1}^n \left[x_j, L\right]L_{mj}&=&\frac{-i\hbar}{2} x_m\left(\sum_{j,k}L_{jk}\gamma_j\gamma_k\right)\nonumber \\
&=&i\hbar x_m L-\frac{i\hbar^2(n-2)}{2}x_m\nonumber.\eea
Leaving the final expression (\ref{RmL})
We are now in a position to prove asserted ladder operations as given in Theorem \ref{Lpmt}.

{\bf Proof of Thm \ref{Lpmt}:} By Lemmas \ref{xmLLemma} and \ref{RmLlemma}, the commutators are 
\bea\fl  [L_m^+,L]&=&\frac{\hbar(n-2)}{2}[x_m,L]-i\hbar R_m-i\sum_{j\ne m} x_j[L_{mj}, L] \nonumber\\
\fl &&+i\left(i\hbar x_m L-\frac{i\hbar^2(n-2)}{2}x_m+\sum_{j=1}^nx_j[L_{mj}, L]\right)-\frac{\hbar(n-2)}{2}[x_m,L]\nonumber\\
\fl &=&-\hbar L_m^{+}. \nonumber \eea

\noindent The above relations can be straightforwardly extended also to $L_k^{(-)}$ by considering its adjoint property.

\section*{References}
\bibliography{bib}
\bibliographystyle{unsrt}
\end{document}